\documentclass[numberedappendix,apj,twocolumn]{emulateapj}

\usepackage[bookmarks,bookmarksnumbered,colorlinks=true, citecolor=blue, linkcolor=black,breaklinks]{hyperref}

\shorttitle{The Hubble Deep UV Legacy Survey}
\shortauthors{Oesch et al.}

\begin{document}

\title{HDUV: The Hubble Deep UV Legacy Survey
\altaffilmark{1}}

\altaffiltext{1}{Based on observations made with the NASA/ESA Hubble Space Telescope, obtained from the data archive at the Space Telescope Science Institute. STScI is operated by the Association of Universities for Research in Astronomy, Inc. under NASA contract NAS 5-26555. All of the HDUV data products are available at the Mikulski Archive for Space Telescopes (MAST) as a High Level Science Product (\dataset[doi:10.17909/T90T2N]{http://dx.doi.org/10.17909/T90T2N}) at \url{https://archive.stsci.edu/prepds/hduv/}.}

\author{P. A. Oesch\altaffilmark{2,3}, 
M. Montes\altaffilmark{4,3},
N. Reddy\altaffilmark{5,$\dagger$},
R. J. Bouwens\altaffilmark{6}, 
G. D. Illingworth\altaffilmark{7},  
D. Magee\altaffilmark{7},
H. Atek\altaffilmark{8}, \\
C. M. Carollo\altaffilmark{9}, 
A. Cibinel\altaffilmark{10},
M. Franx\altaffilmark{6}, 
B. Holden\altaffilmark{7},
I. Labb\'{e}\altaffilmark{6,11}, 
E. J. Nelson\altaffilmark{12}, 
C. C. Steidel\altaffilmark{13}, \\
P. G. van Dokkum\altaffilmark{3}, 
L. Morselli\altaffilmark{14},
R. P. Naidu\altaffilmark{12}, 
S. Wilkins\altaffilmark{10}
}

\altaffiltext{2}{Department of Astronomy, University of Geneva, Ch. des Maillettes 51, 1290 Versoix, Switzerland; \email{pascal.oesch@unige.ch}}
\altaffiltext{3}{Astronomy Department and Yale Center for Astronomy and Astrophysics, Yale University, New Haven, CT 06511, USA}
\altaffiltext{4}{School of Physics, University of New South Wales, NSW 2052, Australia}
\altaffiltext{5}{University of California, Riverside, 900 University Ave, Riverside, CA 92507, USA}
\altaffiltext{6}{Leiden Observatory, Leiden University, NL-2300 RA Leiden, Netherlands}
\altaffiltext{7}{UCO/Lick Observatory, University of California, Santa Cruz, CA 95064, USA}
\altaffiltext{8}{Institut d'astrophysique de Paris, CNRS UMR7095, Sorbonne
Universite, 98bis Boulevard Arago, F-75014 Paris, France}
\altaffiltext{9}{Institute for Astronomy, ETH Zurich, 8092 Zurich, Switzerland}
\altaffiltext{10}{Astronomy Centre, Department of Physics and Astronomy, University of Sussex, Brighton, BN1 9QH, UK}
\altaffiltext{11}{Swinburne University of Technology, Hawthorn, VIC 3122, Australia}
\altaffiltext{12}{Harvard-Smithsonian Center for Astrophysics, 60 Garden Street, Cambridge MA, 02138, USA}
\altaffiltext{13}{Cahill Center for Astronomy and Astrophysics, California Institute of Technology, MS 249-17, Pasadena, CA 91125, USA}
\altaffiltext{14}{Excellence Cluster Universe, Boltzmannstr. 2, D-85748 Garching bei M\"{u}nchen, Germany}
\altaffiltext{$\dagger$}{Alfred P. Sloan Foundation Fellow}

\begin{abstract} 
We present the \textit{Hubble} Deep UV Legacy Survey (HDUV), a 132 orbit imaging program with the WFC3/UVIS camera onboard the Hubble Space Telescope (HST). The HDUV extends and builds on the few previous UV imaging surveys in the two GOODS/CANDELS-Deep fields to provide deep images over a total area of $\sim100$ arcmin$^2$ in the two filters F275W and F336W.  Our release also includes all the F275W imaging data taken by the CANDELS survey, which were aligned using a novel approach and combined with the HDUV survey data. By reaching depths of 27.5-28.0 mag ($5\sigma$ in 0\farcs4 apertures), these are the deepest high-resolution UV data over such a large area taken to date.
Such unique UV imaging enables a wide range of science by the community. Among the main goals of the HDUV survey are: (1) provide a complete sample of faint star-forming galaxies at $z\sim1-3$, (2)  constrain the ionizing photon escape fraction from galaxies at $z\sim2-3$, and (3) track the build-up of bulges and the disappearance of clumpy disk galaxies through reliable internal stellar population properties at sub-kpc resolution out to $z\sim3$. 
The addition of the HDUV data further enhances the legacy value of the two GOODS/CANDELS-Deep fields, which now include deep 11-band $HST$ imaging as well as very deep ancillary data from X-ray to radio, enabling unique multi-wavelength studies.
Here, we provide an overview of the survey design, describe the data reduction, and highlight a few basic analyses on the images which are released to the community as high level science products via the Mikulski Archive for Space Telescopes (MAST).
\end{abstract}


\section{Introduction}

The redshift range $z\sim1-3$, about 2-6 Gyr after the first galaxies formed, straddles the peak of the star-formation (SF) history in the universe \citep{Madau14}. This was a time of critical changes in the galaxy population, as the first passive galaxies emerged and the morphological Hubble sequence started to form \citep[e.g.][]{Conselice14}. 
Over the past years, our understanding of galaxy build-up at $z\sim2$ has benefited from studies at multiple wavelengths from the rest-frame UV through submm/radio. However, these studies have primarily focused on the bright, super-L$*$ population, and the UV-faint population at $z\sim1-3$ whose progenitors provide the bulk of the star-formation rate and stellar mass densities at $z>3$ remains largely unexplored. 

In particular, one major shortcoming in our understanding of faint $z>2$ galaxies is that their star-formation and dust measurements largely rely on UV-based tracers, as only the brighter galaxies are directly detectable through their dust emission in the infrared \citep[e.g.][]{Reddy10}. To better understand faint galaxies at $z>2$, we require detailed observations of their analogs at lower redshifts ($z\sim1-3$), where galaxies are amenable to multi-wavelength study. The low-luminosity galaxies at $z<2$ thus provide an ideal benchmark for interpreting the very high-redshift universe.

Exploring these UV faint, intermediate redshift sources requires deep imaging across the observed 2000-3000 \AA\ range. Previous surveys performed with GALEX or with the Swift Ultraviolet/Optical Telescope are not deep enough \citep[reaching $<25.5$ mag; e.g.][]{Zamojski07,Hoversten09} and have only very coarse spatial resolution ($>2.5 \arcsec$), and are thus not suitable for studies of the low star-formation galaxy population at $z\sim1-3$. However, with the installation of the WFC3/UVIS camera \citep{Kimble08} on the \textit{Hubble} Space Telescope ($HST$) further progress is now possible. 

So far, only a few WFC3/UVIS imaging surveys have been conducted over extragalactic fields. The first was part of the Early Release Science (ERS) program of the WFC3 camera \citep[][]{Windhorst11}. While these data only have a depth of 1-2 orbits, the ERS covered $\sim50$ arcmin$^2$ and could be used for the first constraints on the UV luminosity function of Lyman Break selected galaxies (LBGs) at $z\sim2-3$ \citep[][]{Oesch10d,Hathi10}. Subsequent surveys were deeper, single pointings with WFC3/UVIS over the Hubble Ultra-Deep Field \citep[UVUDF; GO12534, PI: Teplitz;][]{Teplitz13,Rafelski15} and over lensing clusters \citep[GO12201/12931/13389/14209, PI: Siana;][]{Alavi14,Alavi16}.
Only few UV images of extragalactic fields have been released as high-level science products (HLSP) to the community, however, limiting the impact of the WFC3/UVIS camera on extragalactic science.

To obtain a complete census of faint, star-forming galaxies with significant samples that cover a range in luminosity requires deep data spanning a much wider area than covered by previous surveys. This is the primary aim of the Hubble Deep UV Legacy Survey (HDUV), a 132-orbit program which is presented in this paper (GO13872, PI: Oesch). Section \ref{sec:obs} provides the survey overview. In Section \ref{sec:datared}, we describe the data reduction steps, before we present the legacy data that are released to the community together with some basic analysis of the images in Section \ref{sec:dataprod}. Finally, we end with a summary.

Throughout this paper, we adopt a standard cosmology with $\Omega_M=0.3, \Omega_\Lambda=0.7, H_0=70$ kms$^{-1}$Mpc$^{-1}$, i.e., $h=0.7$, largely consistent with the most recent measurements from Planck \citep{Planck2015}. Magnitudes are given in the AB system \citep{Oke83}.

\section{Primary Science Goals}
\label{sec:science}
While this paper is focused on the HDUV data reduction and release, we first provide a short overview of the primary science goals of the survey, which motivated its choice of filters and layout (see Figures \ref{fig:filters} and \ref{fig:layout}). 

\subsection{Low-Luminosity Galaxies at Intermediate Redshifts}
Despite the advent of ALMA, most of our estimates of the star-formation rates at $z>2$ largely rely on UV-based tracers that are calibrated at $z\sim0$. An accurate dust correction is thus crucial for a reliable estimate of the cosmic SFR density. By sampling the rest-frame UV for sources at $z\sim0.5-1.5$, the WFC3/UVIS data from the HDUV survey allow us to select lower-redshift UV faint galaxies in the same light as the high-redshift population, with the advantage that we have additional access to a much longer wavelength baseline, extending from X-rays to the far-IR. These data allow us to measure the dust extinction in the $z\sim1-2$ counterparts of the low-luminosity galaxies that dominate the SFR density at high redshifts. The combination of 10-band HST photometry with spectroscopic redshifts and Balmer decrement measurements enables direct tests for the validity of the local dust attenuation curve \citep[e.g.][]{Meurer99,Calzetti00} as a function of various quantities of interest (e.g. luminosity, mass, age, SSFR). A first analysis of the IRX-$\beta$ relation based on the HDUV survey data is published in \citet{Reddy18}.

\subsection{Ionizing Photon Escape Fractions}
One of the most important questions in observational cosmology today concerns the sources that reionized the universe. Although galaxies during the reionization era at $z\sim6-10$ have been directly detected \citep[e.g.][]{Bouwens15aLF,Finkelstein15,Oesch18}, the answer to this question hinges upon the poorly known escape fraction of ionizing photons ($f_{esc}$) into the inter-galactic medium. Standard reionization calculations conclude that an escape fraction of $f_{esc}=10-20$\% is required for galaxies to drive reionization, consistent with the optical depth measurements by the cosmic microwave background \citep[e.g.][]{Bouwens12b,Robertson15,Greig17}.

However, only very few galaxies with such high escape fractions have been found to date as direct observations of the Lyman continuum (at $\lambda_{rest}<912$\AA) are very challenging \citep[e.g.][]{Vanzella16,Izotov16}. Using the two filters F275W and F336W, the HDUV survey can directly probe the Lyman continuum of star-forming galaxies at $z>2.4$ and $z>3.1$, respectively. However, for a reliable measurement of the escape fraction, deep imaging of large galaxy samples at these redshifts are required to reach the LyC and to average out over the large fluctuations in IGM absorption along different lines-of-sight \citep[e.g.][]{Inoue14}. This motivated the HDUV area and depth.
A first set of LyC emitter candidates at $z\sim2$ has been published in \citet{Naidu17}, and a first analysis of mean escape fractions of strong line emitters can be found in \citet{Rutkowski17}. Further papers are in preparation.

\begin{figure}[tbp]
	\centering
	\includegraphics[width=0.99\linewidth]{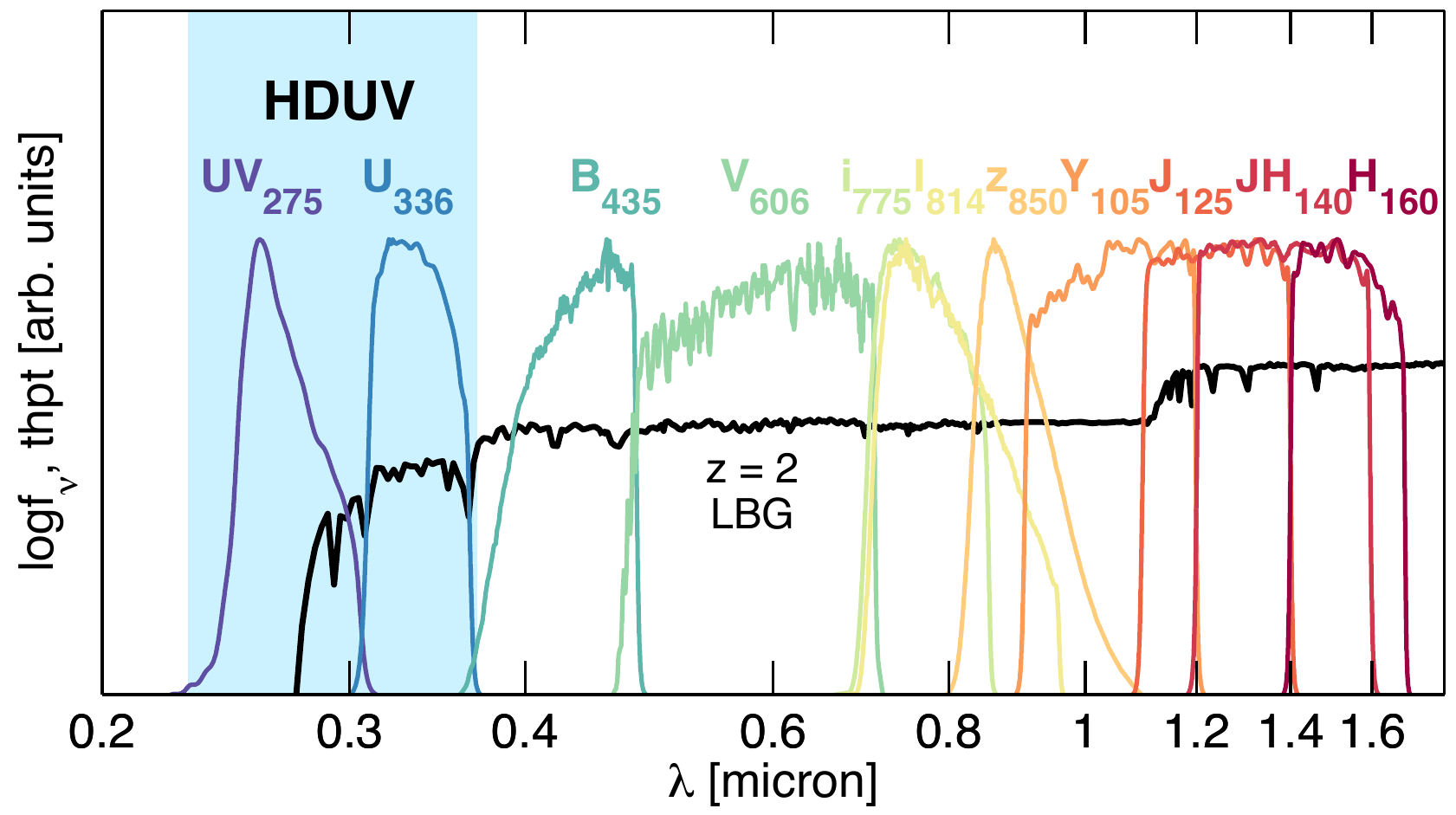}
  \caption{The $HST$ imaging filter set in the two HDUV survey fields. Combining the data of the two workhorse cameras on $HST$, WFC3 and ACS, these fields are now covered with 11-band data spanning a wavelength range from $\sim0.25-1.7$~$\mu$m at excellent spatial resolution of 100-160 mas. The black line represents the spectrum of a typical Lyman break galaxy at $z=2$, which can be selected thanks to the HDUV data.
   }
	\label{fig:filters}
\end{figure}

\subsection{Morphological Evolution at $z<2$}
One of the main open questions in galaxy evolution concerns the physical processes that drive the striking morphological transformation after the peak of cosmic star-formation to form the present-day Hubble sequence \citep[see][for a recent review]{Conselice14}.
By adding deep UV (and parallel optical) HST imaging to the CANDELS Deep data, the HDUV provides resolved, internal multi-color information to faint surface brightness levels at scales of $\sim500$ pc at $z\sim1$. In particular, the combination of UV imaging with WFC3/IR data breaks the dust-age degeneracy and enables a measurement of the age distribution in bulges, clumps, and disks and reveals the sites of recent star-formation in seemingly quiescent ellipticals.
A first set of papers on resolved stellar population properties exploiting the HDUV dataset are published in \citet[][]{Nelson18} and Morselli et al. (2018, in prep).

\begin{figure}[tbp]
	\centering
	\includegraphics[width=\linewidth]{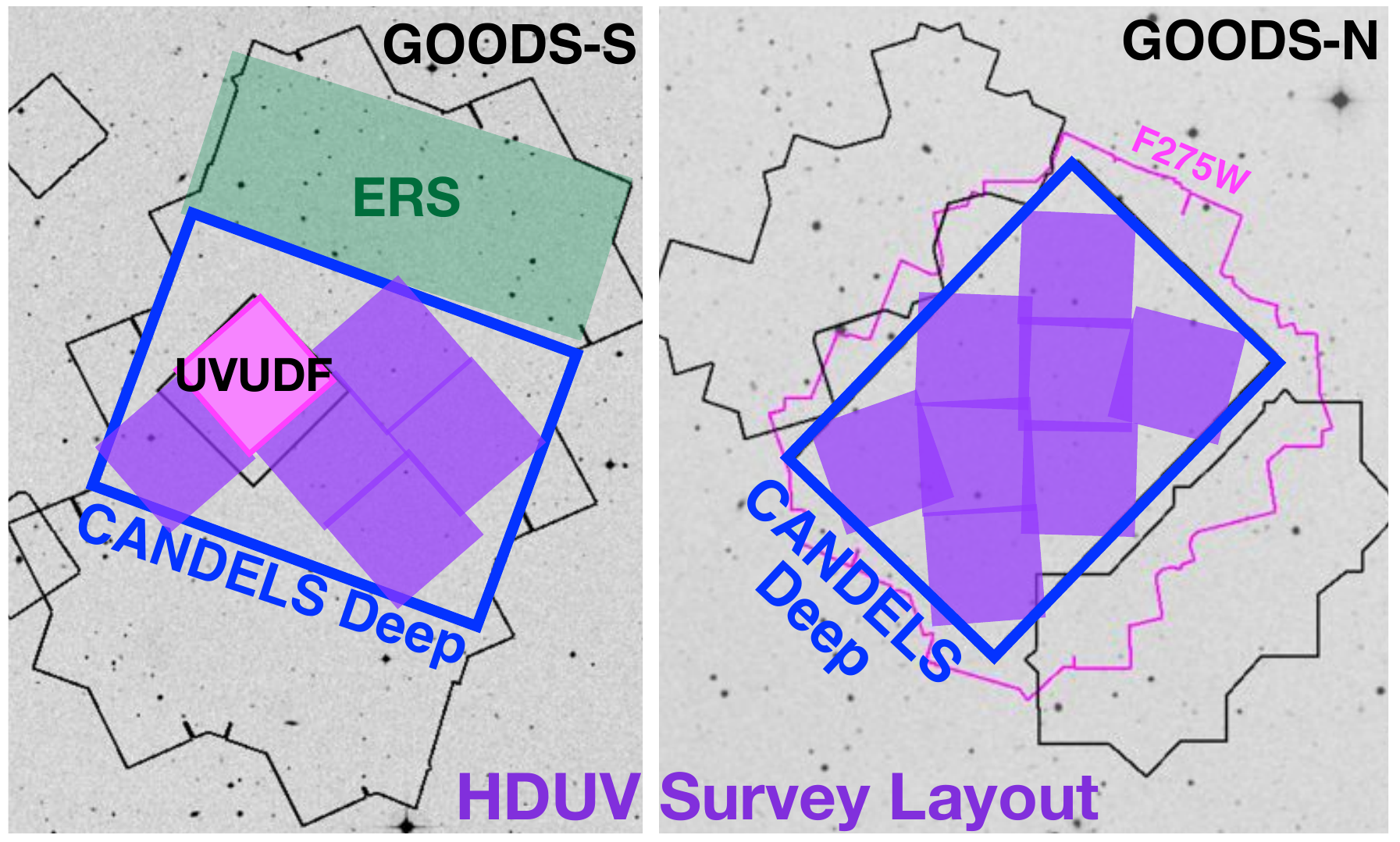}
  \caption{Layout of the HDUV Survey with respect to previous datasets in the GOODS-South (left) and -North (right) fields. The HDUV was designed to optimally build on and extend the pre-existing data to provide deep F275W and F336W imaging over most of the CANDELS-Deep area (blue outline).  In GOODS-S, the HDUV covers 5 pointings around the UVUDF (pink). A larger area (8 pointings) could be covered in GOODS-N making use of the CVZ and building on the limited, pre-existing F275W imaging from the CANDELS survey. When combined with the UVUDF, the total area of WFC3/UVIS imaging in these two fields spans $\sim100$ arcmin$^2$ to a depth of 10 and 8 orbits or deeper in the two filters, respectively, reaching 27.5-28.0 mag at 5$\sigma$. The data release described here includes the F275W images from the CANDELS survey in GOODS-N.}
	\label{fig:layout}
\end{figure}

\section{Survey Design}
\label{sec:obs}

The main goal of the HDUV is to obtain a complete census of faint, star-forming galaxies at the peak of cosmic star-formation, $z\sim1-3$. The HDUV survey is designed to efficiently complement and build on existing, previous datasets. In particular, the HDUV is placed in the GOODS/CANDELS-Deep regions \citep[see][]{Grogin11,Koekemoer11}, which have the best multi-wavelength coverage of all extragalactic fields that are not limited to a single pointing (the latter including the HUDF and the Hubble Frontier Fields, which are covered by different UV programs).

\subsection{WFC3/UVIS Data over the CANDELS Deep Fields}
\label{sec:uvdata}

The primary observations of the HDUV survey include 10-orbit deep F275W and 8-orbit deep F336W filter images over a total of 13 WFC3/UVIS pointings split over the central parts of GOODS-N and S for a total coverage of $\sim100$ arcmin$^2$. In the following, we separately describe these two fields, which have slightly different survey strategies due to the use of the continuous viewing zone (CVZ) in GOODS-N.

\subsubsection{GOODS-S Pointings}
The UVUDF survey \citep{Teplitz13,Rafelski15} already obtained relatively deep UV imaging in three filters F225W, F275W, and F336W for one pointing in the HUDF with an exposure time of 16, 16, and 14 orbits in these filters, respectively (including only post-flash exposures).  Our HDUV program is designed to optimally build around the UVUDF pointing and to cover as much of the CANDELS-Deep region in the GOODS-South field as possible. As shown in Fig \ref{fig:layout}, the HDUV survey obtained five additional pointings around the UVUDF to provide contiguous, deep ($>8$ orbits) coverage over a total of six WFC3/UVIS pointings. This also complements the much shallower, 1-2 orbit ERS imaging in the northern part of the GOODS-South field \citep{Windhorst11}. When combined with all the ancillary $HST$ data, the HDUV survey fields are covered by deep imaging in 11 broad-band $HST$ filters covering a wavelength range of $0.25-1.7$~$\mu$m (Fig \ref{fig:filters}).

The HDUV imaging in the GOODS-S field was obtained at two epochs using two different orientation angles. The first was set close to that of the UVUDF for an optimal, contiguous mosaic. The other orientation was rotated by $\sim90$ degrees, allowing us to test the impact of inefficient charge transfer in the UVIS detectors as described in Section \ref{sec:CTEcorr}.

\subsubsection{GOODS-N Pointings}
The GOODS-North field lies in the continuous viewing zone of $HST$. Due to the low sky background at UV wavelengths, this means that observations can effectively be obtained during the full 96 min orbit of $HST$ without occultation by Earth. By exploiting the CVZ, the HDUV survey speed in GOODS-N was thus boosted by a factor of $\sim2$, allowing us to survey a larger area of 8 pointings. The use of the CVZ, however, prevented us from obtaining observations at an orientation rotated by 90 degrees as was done in GOODS-S.

The CANDELS survey had already obtained a significant amount of UV F275W imaging data in the GOODS-North field during the bright part of its CVZ orbits (GO programs 12443, 12444, and 12445). The HDUV was designed to build upon these existing data in the archive and only obtained 4 orbits of F275W imaging per pointing to reach the same depth as achieved in the GOODS-S field.
The released images of our HDUV survey described in section \S4 include the full CANDELS F275W data, which cover a somewhat larger area than the HDUV itself (see also Table \ref{tab:hduvsummary}). We only include the CANDELS WFC3/UVIS images taken with post-flash and without rebinned pixels, as the other data cannot be calibrated properly and are not photometric due to irrecoverable charge transfer inefficiencies (see Section \ref{sec:CTEcorr}). This means that we only include CANDELS F275W data from programs 12444 and 12445, which were taken between 2012 Sept and 2013 Aug.

\subsubsection{Post-Flash and Full Orbit Integrations}

During the last few years in space, the WFC3/UVIS detectors have degraded, resulting in significant charge transfer efficiency (CTE) issues in low-background observations as we perform here. 
The severity of the CTE issues were first recognized during the UVUDF campaign. As outlined in \citet{Teplitz13}, the CTE degradation leads to data that are not photometric and are thus not usable for science. 

To mitigate these effects, the HDUV WFC3/UVIS images were all obtained with post-flash. In particular, we followed the guidelines by STScI to ensure a background of at least 12 electrons per pixel before reading out the exposure (see also section \ref{sec:background}). Due to this artificial background, it is beneficial to expose each image as long as possible. We thus obtained full-orbit exposures ($\sim2500$~s) for all UV images of the HDUV survey. We further ensured that each pointing was covered by at least 8 exposures in order to be able to reliably mask the higher incidence of cosmic rays due to the long exposure times. This same strategy had previously been adopted for the WFC3/UVIS imaging over the HFFs (PI:Siana).

\subsubsection{Data Acquisition}

The 132 orbits of HDUV data were obtained in 42 independent visits between 2014 Nov 25 and 2015 Nov 16. In GOODS-S these visits were composed of 4 or 5 full-orbit exposures in either F275W or F336W. The exposure time of these images was $\sim2600-2800$ s. In GOODS-N, all visits made use of the CVZ. Thus to reach an equivalent 4-orbit exposure, only 2 actual orbits are required for each filter. Since the CANDELS survey already obtained $\sim6$-orbit deep F275W imaging over GOODS-N, the HDUV obtained only one 2-orbit visit in F275W (with 4 exposures of $\sim2400$ s) and two 2-orbit visits in F336W (with 4 exposures each).
This strategy ensures efficient and accurate alignment of the HDUV data by first stacking all the images in the same visit to remove cosmic rays and reveal a sufficient number of sources at high signal-to-noise (see Section \ref{sec:alignment}). 

The dither pattern adopted for the HDUV exposures varied for different visits, but was a combination of the standard WFC3-UVIS-DITHER-BOX and WFC3-UVIS-DITHER-LINE. These allow for an optimal sampling of the PSF given the number of exposures in a given visit. Additionally we made sure that each visit contained a large scale dither which steps across the inter-chip gap (using POS-TARG offsets of $\sim$4 arcsec or WFC3-UVIS-GAP-LINE). The latter also ensured that different columns were separated sufficiently for removal of hot columns and a blotchy pattern in the background that is impossible to remove in data without large scale dithers (e.g., epoch 3 of the UVUDF dataset).

\subsection{B-band Parallel Imaging Data}
\label{sec:bdata}

In addition to the primary data, the HDUV program obtained ACS F435W ($B_{435}$) imaging in parallel to the WFC3/UVIS exposures.
These exposures provide important additional depth at blue wavelengths over these legacy fields, in preparation for the James Webb Space Telescope (JWST), which will not be sensitive at $\lambda\lesssim6000$ \AA. In particular, the HDUV added $\sim4$ orbit deep F435W imaging data over 10 ACS pointings in the extended Chandra Deep Field South, and $\sim6$ orbit-depth data over 8 pointings around the GOODS-N field. This parallel $B_{435}$-band imaging strategy has already been employed by several other recent WFC3 surveys over extragalactic legacy fields, e.g. the HUDF09 \citep{Oesch10a,Bouwens10a}, HUDF12 \citep{Ellis13}, CANDELS \citep{Grogin11,Koekemoer11}, or the UVUDF \citep{Teplitz13}. 

We will not describe the $B_{435}$ data in detail in this paper. For the GOODS-South field, the first ACS images (taken before Sept 2015) have already been fully reduced and incorporated in the most recent release of the Hubble Legacy Field (HLF) South by our team \citep{Illingworth16}, which combined all available ACS and WFC3/IR data in that area of the sky\footnote{\url{https://archive.stsci.edu/prepds/hlf/}}. 
The ACS data in GOODS-N are being processed and reduced with the same tools as used for the HLF and will be distributed to the community in a future release.

\section{Data Calibration and Reduction}
\label{sec:datared}
In this section we describe the data reduction process for the WFC3/UVIS images. A special pipeline for the reduction of these data was written based on previous WFC3 and ACS data reduction pipelines developed by our team and used, e.g., for the public image release of the Extreme Deep Field (XDF) and the Hubble Legacy Field (HLF) \citep{Illingworth13,Illingworth16}.

The pipeline is written in python and makes heavy use of the \texttt{drizzlepac} software package developed by STScI\footnote{\url{http://drizzlepac.stsci.edu/}}, which includes the basic processing steps required for $HST$ data reduction.
The first step is to download all individual raw frames from the MAST archive\footnote{\url{https://archive.stsci.edu/}}. The raw frames are then directly corrected for CTE losses before application of an initial cosmic ray identification for each individual image, subtraction of the overall background as well as a sky dark, and manual masking of satellite trails. The main steps are described in more detail in the next subsections.

\subsection{CTE Correction}
\label{sec:CTEcorr}
We adopt the pixel-based CTE correction for WFC3/UVIS images provided by STScI\footnote{\url{http://www.stsci.edu/hst/wfc3/tools/cte_tools}}. As of February 2016, this correction is part of the standard data processing performed by STScI when requesting images from the archive. Since our data were processed before that date, we performed these corrections ourselves using the same code.

The CTE correction method is based on an empirical model of hot pixels in dark exposures, and it can be used to correct the fluxes of sources and to restore their morphology. Note, however, that the accuracy of any correction method is limited. In particular, one cannot restore the somewhat reduced signal-to-noise ratio due to smearing of galaxy flux, which is subject to the readout and background noise of a larger number of pixels compared to ideal observations without charge transfer losses.

\begin{figure}[tbp]
	\centering
	\includegraphics[width=\linewidth]{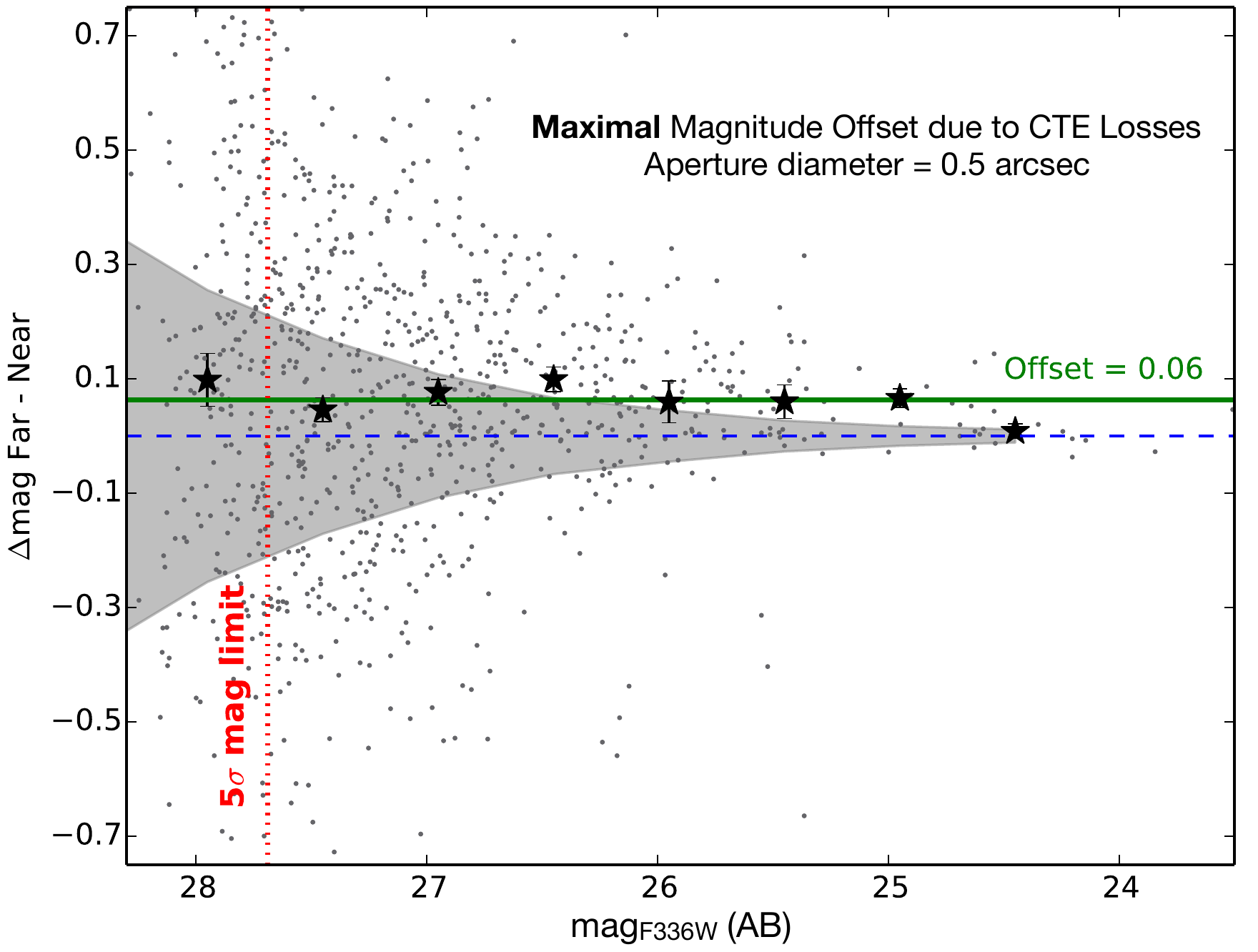}
  \caption{Measured magnitude differences of galaxies within 0\farcs5 apertures when observed close to the readout (i.e., with few charge transfers, $\lesssim600$) compared to when observed far from readout (i.e., with many charge transfers, $\gtrsim1500$) as a function of total F336W magnitude. Small dots correspond to individual measurements, while stars represent binned median values. The difference is measured in individual CTE-corrected stacks of epoch 1 and epoch 2 data in the GOODS-S field, which were obtained with a $\sim90$ deg offset in their position angle. The gray curve represents the expected scatter purely due to a decreasing signal-to-noise at fainter magnitude. The CTE correction works very well, resulting only in a median residual flux loss of $\sim0.06$ mag and in an additional scatter of $\sim0.1$ mag, which appears to be independent of magnitude. Given that these galaxies were selected to be the extreme cases, with the most charge transfers, the average galaxy in the field is expected to show a magnitude bias of less than 5\%.}
	\label{fig:CTEtest}
\end{figure}

Within our dataset, we can directly test the accuracy of the CTE correction for extragalactic fields, and we find residual biases of less than $\sim6\%$.
In particular, in GOODS-S, we obtained the UV imaging in two separate epochs with position angles separated by about 90 degrees. We can thus select sources which are close to the readout in one epoch ($<600$ charge transfers), but are far from readout (i.e., undergo many charge transfers, $>1500$) in the other epoch. This flux comparison in the GOODS-S F336W image is shown in Figure \ref{fig:CTEtest}. As can be seen, the median CTE losses after correction are nearly independent of magnitude at $\sim6\%$, but they do add an additional $\sim$10\% to the flux uncertainty. Note, however, that this is the maximal effect for the sources undergoing the largest number of charge transfers. The average galaxy flux in our and other recent extragalactic WFC3/UVIS surveys taken with postflash exposures will thus only be affected at the $<5\%$ level after CTE correction.

\subsection{Cosmic Ray Rejection}

Cosmic rays are a major concern for our WFC3/UVIS images, and they require special attention. Before processing and combining the data further, the next step is therefore an initial cosmic ray identification using Laplacian edge detection \citep{vanDokkum01}. In particular we make use of a python wrapper to \texttt{lacosmic}\footnote{\url{http://obswww.unige.ch/~tewes/cosmics_dot_py/}}. 

This initial cosmic ray mask is required for accurate background subtraction of each frame and improves the initial drizzling of images on a visit-by-visit basis (see next sections). For the final science image, we erase these initial cosmic ray masks, however, and we use the standard \texttt{drizzlepac} routines for identifying cosmic rays in the final drizzle run.

\begin{figure*}[tbp]
	\centering
	\includegraphics[width=0.9\linewidth]{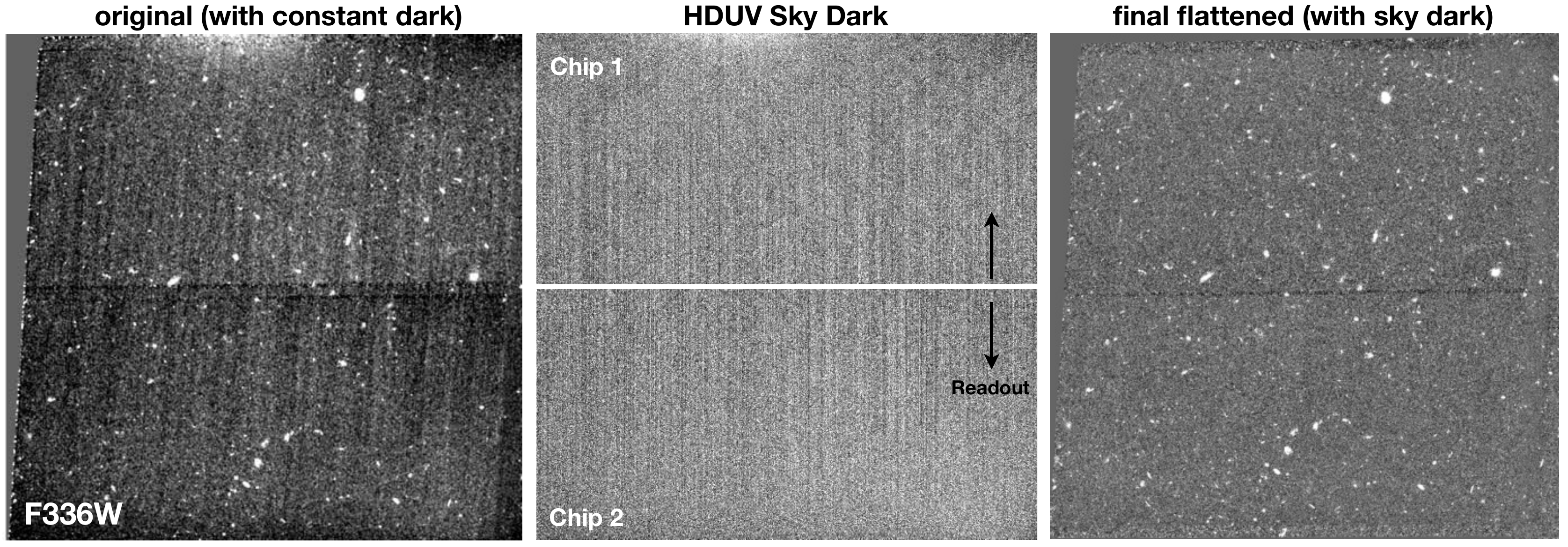}
  \caption{The improvement in image quality thanks to the use of a sky dark image in the HDUV data reduction. The left panel shows a single visit of F336W data drizzled using the standard dark frame. Significant residual structure is clearly visible, including severe striping along the readout direction (i.e. hot and cold columns) as well as a significant glow at the upper edge. These can be largely removed with a sky dark. The middle panel represents the corresponding sky dark image computed for the first epoch data of the HDUV in the F336W filter. 
  Such sky dark images are computed for each filter and for the HDUV and CANDELS data separately. They are then subtracted from each corresponding frame before further processing. The right panel shows the improvement in image quality after application of the sky dark on the same F336W visit as in the left panel.}
	\label{fig:skyDark}
\end{figure*}

\subsection{Background Levels}
\label{sec:background}

The recommended background level for WFC3/UVIS exposures is a minimum of 12 electrons in order to sufficiently fill the charge traps and the derivation of CTE corrected images, which are photometric (see also Section \ref{sec:CTEcorr}). Since the UV sky background is lower than required even for full-orbit exposures as adopted here, an artificial background has to be applied, which is done through pre-flashing the WFC3/UVIS detectors (so-called post-flash observations). The HDUV post-flash values were set to reach a theoretical electron level of at least 14 electrons for both filters to account for the variation in the post-flash background across the detector and to ensure that most pixels are indeed exposed to a background of at least 12 electrons. When factoring in the dark current and expected minimum sky background level, the additional requested background electrons from post-flash were set to 10 and 6 for the F275W and F336W images, respectively.

In the GOODS-S images, these settings resulted in the expected range of measured backgrounds in the data ranging from a minimum of 13 to a maximum of 18 electrons, meaning that all the data could be properly CTE corrected.
In the GOODS-N field, we measure similar background levels for the F275W images. However, the use of the CVZ also came with some disadvantages. Unfortunately, scheduling of CVZ orbits did not allow us to identify which part of the orbit would be bright and which would be dark. During the bright part, the background in the F336W can be significantly elevated, while the F275W images are not affected. The backgrounds seen in the F336W images in GOODS-N thus span quite a large range, from the minimum of 12 electrons up to a maximum of 30 electrons per exposure. The mean background in the GOODS-N F336W images was 20.1 electrons, i.e., unfortunately, almost twice the required minimum background value. This was expected, but our team decided that the benefits of using boosted survey efficiency, which allowed us to cover a larger survey area, outweighed the slight decrease in the F336W image depth. As shown in Section \ref{sec:depth}, the final drizzled F336W image of the GOODS-N field is 0.2 mag shallower than the corresponding low-background data in GOODS-S.

\subsection{Sky Darks}

Until recently, the standard dark images provided by STScI consisted of a single, constant dark value together with a hot pixel mask. As pointed out, e.g., by \citet{Rafelski15}, this description of the dark signal is not sufficient for observations at  low background levels such as those seen in the UV, even after the artificial post-flash background. The WFC3/UVIS images show significant structure in their dark images, which clearly shows up when combining images at a depth of several orbits as available here. In particular, the combined WFC3/UVIS images exhibit a significant ``striping" along the readout direction when using the standard dark frames (Fig \ref{fig:skyDark}).

For the UVUDF reduction \citet{Rafelski15}  derived their own dark frames from the STScI calibration files, which they also corrected for CTE issues. This results in much improved and flatter UVIS images. STScI now provides similar dark frames as of early 2016.

For the HDUV we decided to apply a different, and simpler approach, which we found to result in even better and flatter images than simply using the CTE corrected dark frames. Given the large number of exposures at different sky positions obtained for the HDUV, we can compute a sky dark from our exposures themselves. This is simply obtained by masking all the pixels which are known to contain real sources (based on a segmentation map of the $B_{435}$ reference frame), and median stacking all the images of a given filter taken within a given epoch. This is done on the already dark-corrected and background-subtracted frames. 

An example of this sky dark is shown in Figure \ref{fig:skyDark}. Significant structure in the dark frame is evident, including striping and a glow at the upper edge of chip 1. The blotchiness on chip 1 may be related to residual flat fielding uncertainties, as some of its structure appears to be correlated with structure in the flat field.

\begin{figure*}[tbp]
	\centering
	\includegraphics[width=0.95\linewidth]{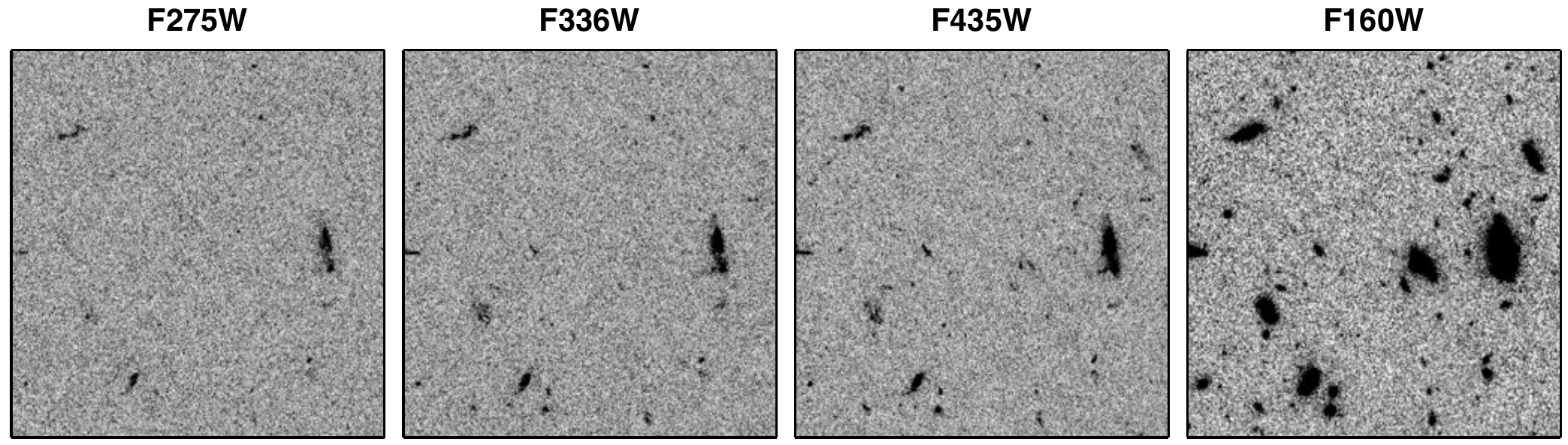}	
  \caption{Examples of images from the HDUV (left two panels; F275W and F336W) and ancillary data (right two panels; F435W and F160W) in the GOODS-South field. The images are 0.5~arcmin on a side. Note the flat background in the WFC3/UVIS data, which was only achieved after subtraction of the sky dark. Clearly, the surface density of sources decreases to shorter wavelengths as the average galaxy exhibits a relatively red color, and sources at $z\gtrsim1.7$ start to disappear from the UV imaging due to the Lyman break (see Fig \ref{fig:filters}). The low number density of UV-bright sources and the very high incidence of cosmic rays make it particularly challenging to align such UV imaging data (see \S \ref{sec:alignment}). }
	\label{fig:imageExamples}
\end{figure*}

Sky darks similar to that shown in Figure \ref{fig:skyDark} are computed for each filter and for each observing epoch. They are then subtracted from each corresponding raw frame, before the final image combination. This approach resulted in flat and clean final drizzled images (see \ref{fig:imageExamples}).

We note that the updated, CTE-corrected dark frames provided by STScI as of early 2016 capture most of the structure seen in our sky darks, including the striping pattern. Therefore, data that is processed with those dark files will look much cleaner than what is shown in the left panel of Figure \ref{fig:skyDark}.

\subsection{Alignment}
\label{sec:alignment}

The high incidence of cosmic rays and the absence of a large number of bright sources in a single exposure makes it very difficult to align UV images of extragalactic fields. While our HDUV survey was designed to facilitate the image alignment within a visit, the previous CANDELS F275W imaging over GOODS-N required special treatment.
Below we describe the two different strategies that were adopted to overcome this problem for the two different datasets combined in the final release image. 

For both fields, the reference frames onto which we match the HDUV reductions were chosen to be the GOODS F435W images. While those images still contain a few residual cosmic rays in the gaps of the ACS chips, the $B_{435}$ band lies closest to our UV images and thus ensures that any possible morphological k-correction of sources is minimal. In particular, we chose the F435W images released by the 3D-HST team\footnote{\url{http://3dhst.research.yale.edu/Data.php}} and drizzled the HDUV images to the same pixels to facilitate an easy use of the HDUV data for multi-wavelength analyses.

\subsubsection{HDUV Alignment}
The HDUV survey was designed to allow for easy image alignment by obtaining several images of the same filter in a given visit.
The relative positions of exposures within a visit are accurate enough that they can easily be combined and cosmic ray cleaned, before a global shift is determined for each visit. The increased depth of the images in an individual visit also ensures that a sufficient number of extragalactic sources is detected to enable a reliable determination of the global shifts.

For each visit, we use \texttt{astrodrizzle} to create a cosmic-ray cleaned, drizzled image of all the images taken in the same filter. \texttt{SExtractor} is then run on this image to identify sources. Residual cosmic rays are cleaned from the resulting catalog based on their location in the magnitude-size diagram. The tool \texttt{superalign}\footnote{\url{https://github.com/dkmagee/superalign}} is then used to match the sources detected in an individual image with sources in our reference frame. \texttt{superalign} returns a matched list of sources in addition to an image shift and rotation for each visit. These shifts are then applied to each input image of a given visit using the astrodrizzle task \texttt{updatehdr.updatewcs\_with\_shift}, which updates the header of each input image with the correct coordinates, matched to the reference frame. These images are then ready to be used for the final drizzle run to create the final release images.

\subsubsection{CANDELS F275W Alignment}
A reduction of the CANDELS F275W imaging in GOODS-N was not publicly available at the time of our analysis. 
These data thus had to be processed as part of the HDUV reduction. The CANDELS UV images are somewhat more complicated to align, since they were taken exclusively in the bright part of each CVZ orbit, with only one F275W image (of $\sim1500$ s) per visit. The visit-by-visit alignment strategy that was adopted for the HDUV data is thus not applicable.

The short exposure time of the CANDELS images does not reveal many real sources in an individual exposure.
Furthermore, the relative coordinates of images taken in different visits are not accurate to the subpixel level.
Thus, it was not possible to align individual F275W images directly. However, the CANDELS survey obtained a short $\sim$400~s long-pass F350LP image in each visit immediately before the F275W exposure. Those images contain enough high-S/N sources for an accurate alignment. 

The strategy was thus to compute relative shifts for each CANDELS visit based on this F350LP image relative to our reference image using the same procedure as described for the HDUV data. These relative shifts were then applied to the F275W image headers again using astrodrizzle tasks. We tested that this strategy results in an accurate alignment of the F275W images both through visual inspection of individually drizzled CANDELS F275W images and through the quantitative comparison of the full stack of CANDELS F275W data relative to the stack of the HDUV F275W data in GOODS-N.

\begin{figure*}[tbp]
	\centering
	\includegraphics[width=0.92\linewidth]{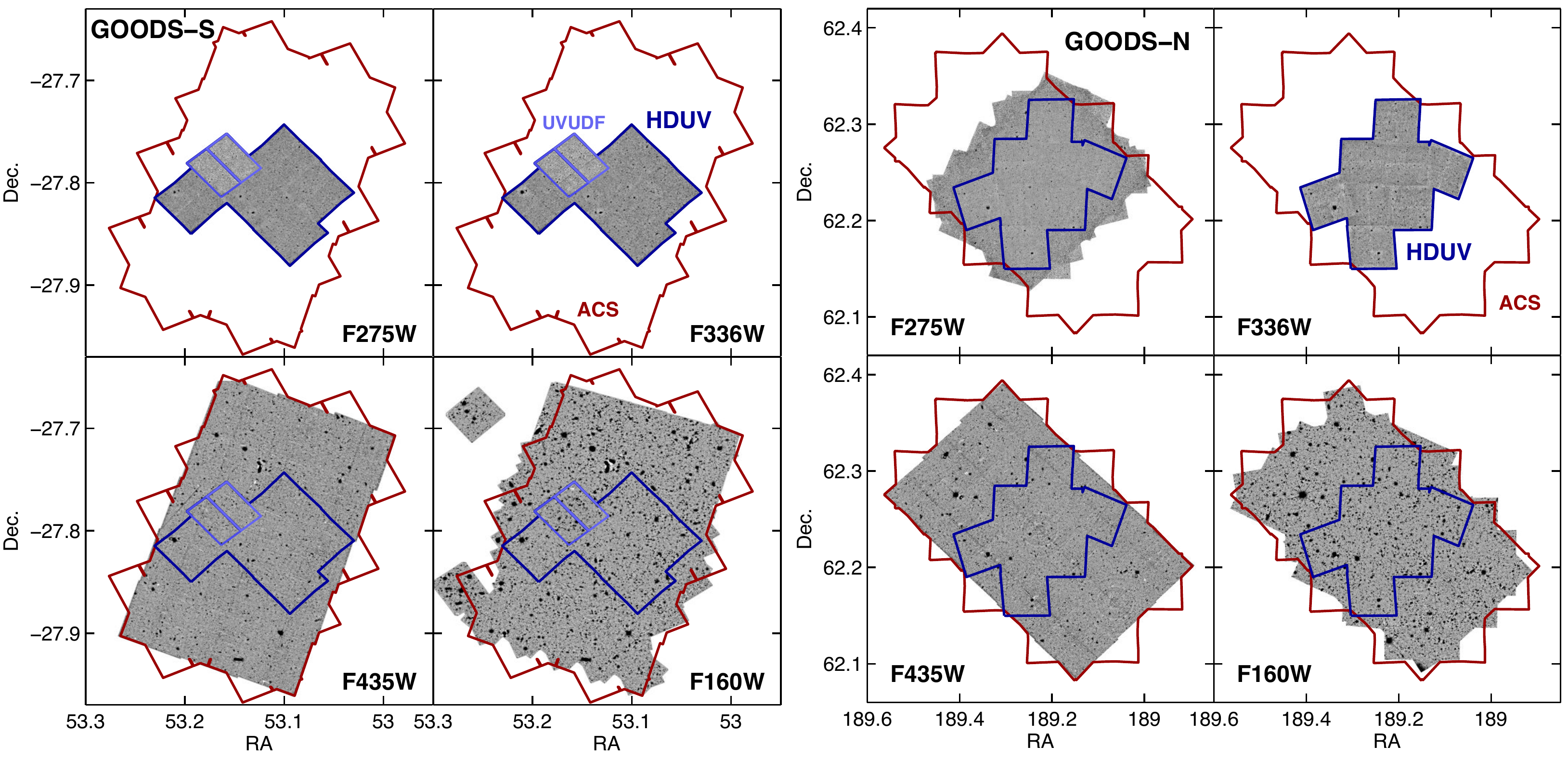}	
  \caption{The HDUV images and their relative location to previous $HST$ data in the GOODS-S (left) and GOODS-N (right) fields. The top panels show the WFC3/UVIS filter data in F275W and F336W of the HDUV processed images. The lower left shows the ACS F435W data from the GOODS survey and the lower right panels show the existing WFC3/IR F160W data. The dark blue outline shows the HDUV footprint. In GOODS-S the HDUV release includes the UVUDF data (pale blue outline). In GOODS-N, the HDUV survey added 4 orbits to the pre-existing F275W imaging data from the CANDELS observations. The latter are also processed and released as part of the HDUV data. Hence the F275W  image in GOODS-N extends over the HDUV survey footprint. The red outline shows the full GOODS ACS coverage. }
	\label{fig:survey}
\end{figure*}

\begin{deluxetable}{lll}
\tablecaption{HDUV Summary \label{tab:hduvsummary}}
\tablewidth{0.85\linewidth}
\tablecolumns{3}
\startdata
\tablehead{\colhead{} & \colhead{HDUV-South} & \colhead{HDUV-North}} 
R.A.  &  03h 32m &   12h 37m \\
Dec.  & $-$27$^\circ$~48\arcmin   &  62$^\circ$~14\arcmin  \\
Pixels  & 17500$\times$19700  &  20480$\times$20480 \\
\cutinhead{F275W} 
Area 	     &  43.4 arcmin$^2$   &  56.5 arcmin$^2$ \\  
 	     &    &  (39.1 arcmin$^2$)\tablenotemark{*} \\  
Exposure Time\tablenotemark{$\dagger$}         &   10.1 orbits  &  10.5 orbits \\
         &     &  (3.8 orbits)\tablenotemark{**} \\
Depth         &   27.6 mag  &  27.4 mag\\
FWHM          &    108 mas      &  96 mas    \\
\cutinhead{F336W} 
Area	         &  43.4 arcmin$^2$   &  56.5 arcmin$^2$   \\
Exposure Time         &   9.1 orbits  &  8.4 orbits \\
Depth        &   28.0 mag   &  27.8 mag \\
FWHM          &   91 mas       &     95 mas 
\enddata
\tablecomments{Depth measurements are averaged over the HDUV survey footprint. They correspond to $5\sigma$ detections measured in circular apertures of 0\farcs4 diameter and are not corrected to total magnitudes. The point-source correction to total is 0.2 mag. Release images include the UVUDF in GOODS-S and CANDELS F275W data in GOODS-N.}
\tablenotetext{$\dagger$}{Average exposure time in orbits, assuming 2400~s per orbit. }
\tablenotetext{*}{Additional area from CANDELS-UV data that is covered by at least two exposures.}
\tablenotetext{**}{Corresponding to area outside of the HDUV footprint.}
\end{deluxetable}

\section{Data Products and Basic Analysis}
\label{sec:dataprod}

The last step of the data reduction is to combine the calibrated WFC3/UVIS images of a given field and taken with a given filter into a final stack. We used the tool \texttt{astrodrizzle} to drizzle these images to the same pixel frame as the GOODS/CANDELS images released by the 3D-HST team\footnote{\url{http://3dhst.research.yale.edu/Data.php}} in order to make multi-wavelength analyses by the community particularly easy. In this last step, we erase the previous cosmic-ray masks of all the input images and use the standard astrodrizzle settings for final cosmic ray identification in the image creation. These final images have sizes of 17500$\times$19700 and 20480$\times$20480 pixels, in GOODS-S and GOODS-N, respectively, with a pixel scale of 60 mas.

For completeness and ease of use, our released image in GOODS-S also includes the v2 image of the UVUDF\footnote{\url{https://archive.stsci.edu/prepds/uvudf/}} \citep{Teplitz13,Rafelski15}. This was matched to the 3D-HST pixel grid using \texttt{swarp} \citep{Bertin02} and added to the HDUV imaging using a weighted mean based on the drizzled weight maps. The full HDUV data release thus includes 6 pointings of WFC3/UVIS images in the GOODS-S and 8 pointings in GOODS-N (see Fig. \ref{fig:survey}) for a total coverage of almost exactly 100 arcmin$^2$.

The version 1 release of the \textit{HDUV} data products is available on the MAST High Level Science Products (HLSP) archive\footnote{\url{https://archive.stsci.edu/prepds/hduv/}}. 
This includes the drizzled science images together with the rms maps. The latter have been derived from the weight images produced by drizzle and have been corrected for correlated noise introduced by the drizzle algorithm using the equations in \citet{Casertano00}.

For convenience, the release webpage also includes a preliminary source catalog of the HDUV footprint with photometry, photometric redshifts, and stellar population properties. These catalogs are very closely related to the ones released by the 3D-HST team \citep{Skelton14}. In particular, they are based on the same near-infrared detection image and aperture photometry was performed in a consistent way to seamlessly integrate the two UV flux measurements from the HDUV survey. For more details on the source catalog, see the readme files and descriptions on the release page.

An additional version of the $HDUV$ images in the GOODS-S field is drizzled to the pixel grid of the Hubble Legacy Field data \citep{Illingworth16}, which we separately make available on that MAST archive page\footnote{\url{https://archive.stsci.edu/prepds/hlf/}}. That image also includes our reduction of the shallower ERS WFC3/UVIS data, which have not previously been made available to the community.

In the following, we present a few basic analyses performed on the release images from the HDUV page.

\begin{figure}[tbp]
	\centering
	\includegraphics[width=\linewidth]{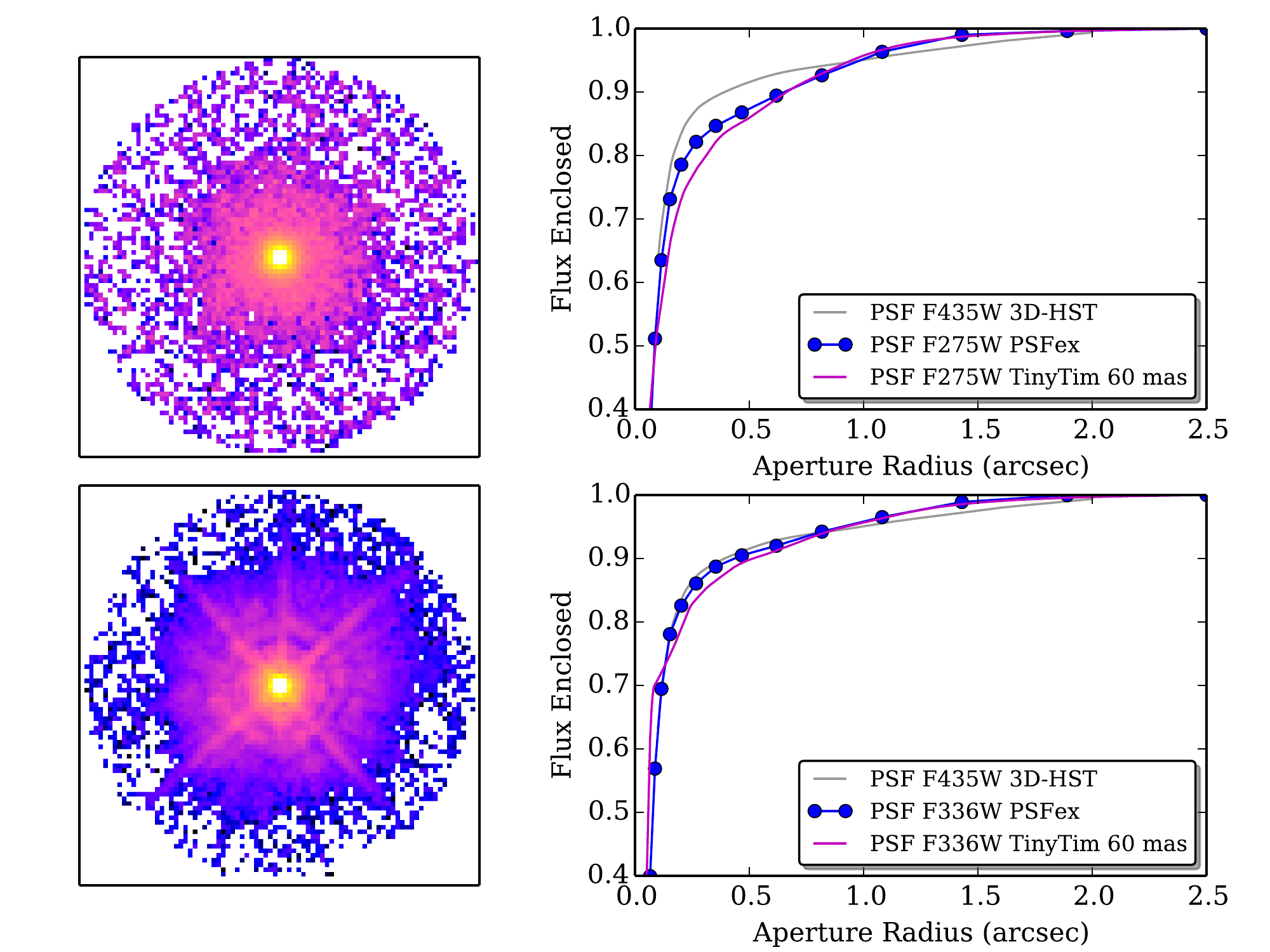}
  \caption{The PSFs of the HDUV data. The left panels show stamps of the F275W (top) and F336W (bottom) PSFs. Images are 5\arcsec on a side. The right panels compare the PSF enclosed flux with the B$_{435}$-band PSF as well as with the theoretically determined TinyTim PSFs of the corresponding WFC3/UVIS filters. The PSFs for the two fields are very consistent with each other.}
	\label{fig:psfcomparison}
\end{figure}

\subsection{Point-Spread Functions}

The point-spread functions (PSFs) of the HDUV images are derived separately for the two fields (GOODS-N and GOODS-S) in the two filters, F275W and F336W, using the software package PSFex \citep{Bertin11}. PSFEx identifies and fits a nth-order polynomial to all the stars in the field-of-view, which is then used to construct the image of the PSF. 
Figure \ref{fig:psfcomparison} shows those images together with the enclosed flux as a function of radius. 

The full-width-at-half-maximum (FWHM) measured for the PSFs are $\sim0.1$\arcsec for both the F275W and F336W images (see Table \ref{tab:hduvsummary}). These are consistent with the theoretically computed PSF widths using TinyTim \citep{Krist11}, and the F336W image PSF is very similar to the PSF in the ACS F435W filter image.
Even though the F275W filter samples shorter wavelengths than F336W, its PSF FWHM is slightly larger, as $HST$ is no longer diffraction limited shortward of the $V$-band due to micro-roughness of the mirror affecting the UV PSF \citep[see also][]{Windhorst11}.

\begin{figure}[tbp]
	\centering
	\includegraphics[width=0.98\linewidth]{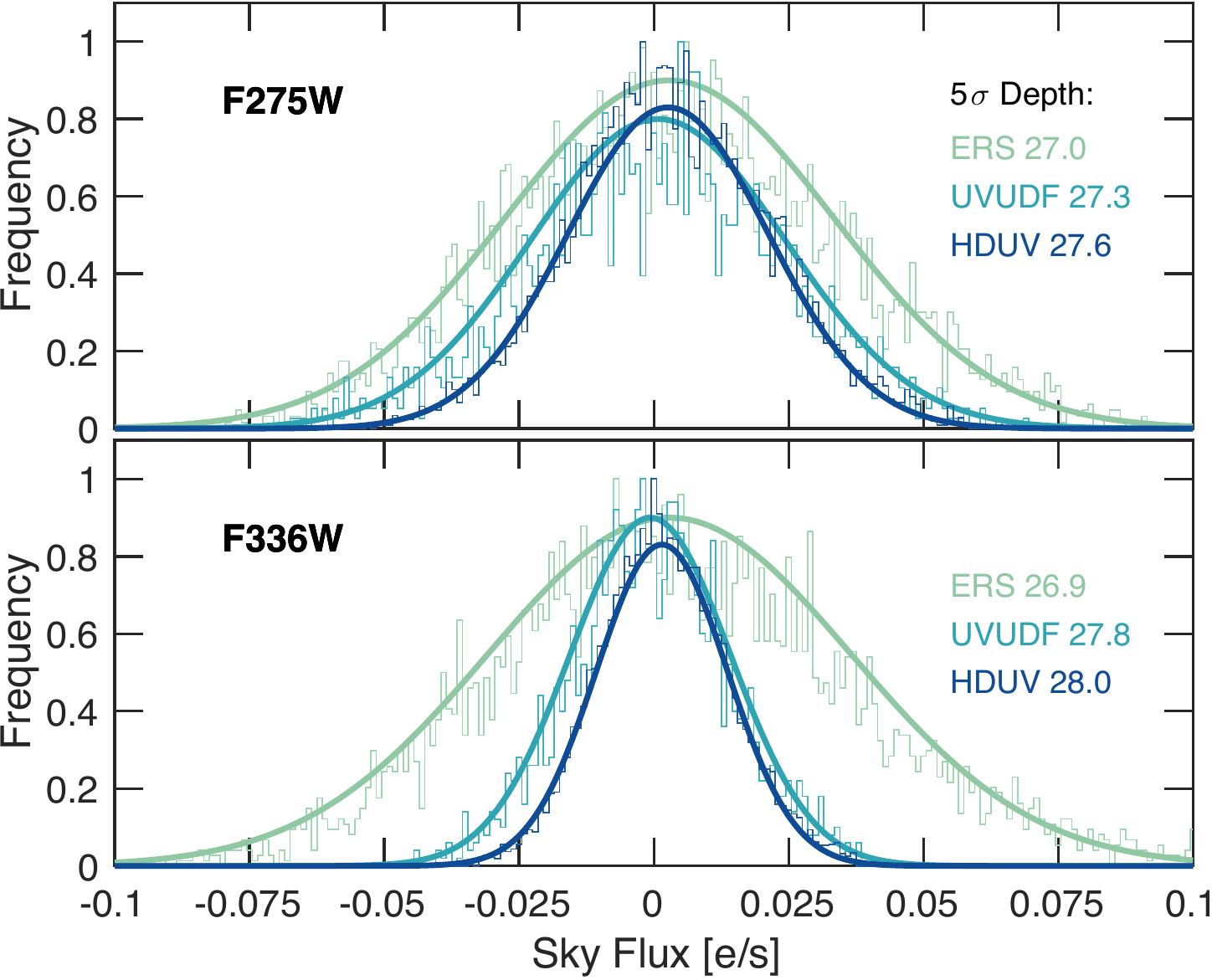}
  \caption{Normalized histograms of sky flux measurements in empty apertures of 0\farcs4 diameter in the GOODS-S field. The top panel shows the measurement for the F275W filter, while the bottom shows F336W fluxes. They have been normalized to a common zeropoint of 25 mag to facilitate the comparison. The different curves represent different surveys. Despite longer exposure time, the UVUDF is slightly shallower than the HDUV data due to problems with uneven background (see text for details). The gains relative to the short ERS exposures of both the HDUV and the UVUDF in the F275W filter are reduced due to the need for added background in post-flash observations.  }
	\label{fig:depths}
\end{figure}

\subsection{Depths}
\label{sec:depth}

The image depths are derived directly from the images by placing several thousand circular apertures in empty sky regions of the HDUV images and measuring their enclosed flux. The dispersion of these flux histograms represents the most accurate, empirical measurement of the average image depth. Within circular apertures of 0\farcs4 diameter, we measure 5$\sigma$ depths of 27.5 and 27.9 in the two filters F275W and F336W, respectively, averaged over the full HDUV survey footprint in GOODS-N and S (see Table \ref{tab:hduvsummary} for the depths split by field). The average correction from these apertures to a total flux is 0.2 mag. The GOODS-S data are deeper by 0.14 and 0.26 mag compared to the GOODS-N data due to the shorter exposure times used in the F275W CANDELS images, due to the higher backgrounds in some of the F336W images during CVZ observations, and due to slightly smaller Galactic dust extinction. The quoted depths are corrected for extinction using the maps from \citet{Schlafly11}, which result in offsets of 0.043 mag and 0.035 mag in GOODS-S, and 0.066 mag and 0.053 mag in GOODS-N, for F275W and F336W, respectively.

Interestingly, performing the exact same measurement on the UVUDF images (v2.0), we measure depths that are 0.2-0.3 mag shallower than the HDUV GOODS-S image (27.3 and 27.8, respectively), despite the longer exposure times of 14-16 orbits (see Figure \ref{fig:depths}). This is due to the uneven background of the UVUDF images, which unfortunately were taken without any large scale dithers in the only usable epoch 3 data. Note that the relatively large discrepancy of the above numbers with the depths presented in Table 1 of \citet{Rafelski15} is due to the way the UVUDF team derived their depths. They compute the single pixel-by-pixel variations in random locations of the image, average those out over the image, and rescale them to 0\farcs4 diameter apertures, only correcting for a theoretical correlation coefficient introduced by the drizzle algorithm. However, this procedure significantly overestimates the actual depth of the UVUDF images (by $\sim0.5$ mag), as it does not account for the blotchy background pattern which is still present in the UVUDF v2 reduction and that is not captured in the theoretical correlation coefficient. A more detailed discussion of this can also be found in \citet{Japelj17}, who find the same depth in circular apertures of the UVUDF F336W as we do here.

The same aperture depth measurement performed on our own reduction of the ERS images\footnote{While we do not release the ERS UV images as HLSPs here, our upcoming v2 release of the Hubble Legacy Field will include them: \url{https://archive.stsci.edu/prepds/hlf/}.},
 results in measured depths of 27.0 and 26.9, respectively. These images were taken soon after commissioning of the WFC3 camera and did not yet require post-flashing the detector to counteract CTE issues. This is particularly noticeable in the F275W filter image, where the required post-flash levels are very high. Thus, despite the significantly longer exposure time ($\sim5\times$), both the HDUV and the UVUDF images are only $\sim0.6$ and $\sim0.3$ mag deeper, respectively, than the ERS F275W data.

\begin{figure}[tbp]
	\centering
	\includegraphics[width=0.98\linewidth]{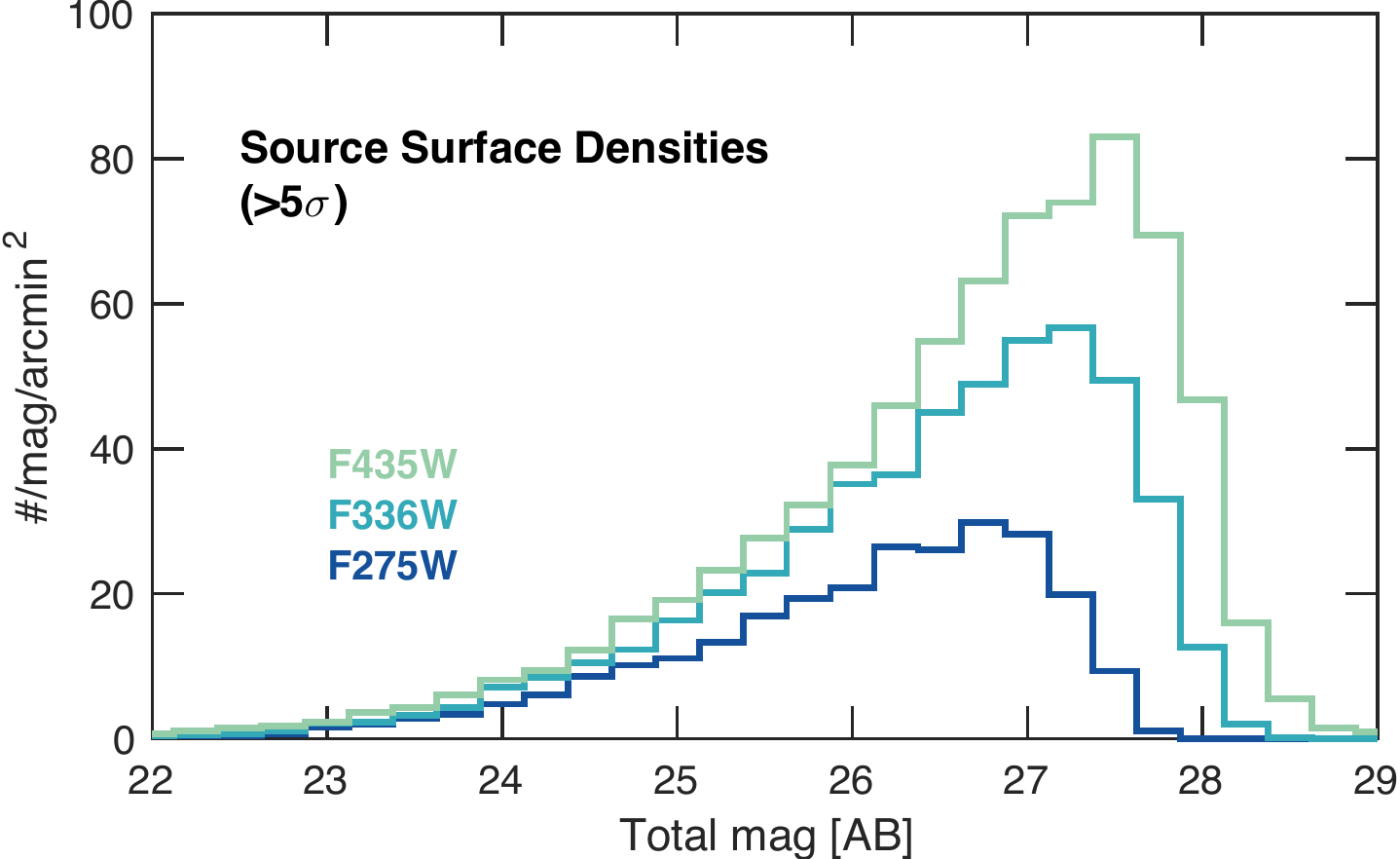}
  \caption{Source surface densities in different bands within the HDUV survey footprint. Shown are the numbers of sources in the ACS F435W band (red), as well as the UVIS F336W and F275W (dark and pale blue, respectively) that are detected at $>5\sigma$ with SExtractor. The depth of the HDUV survey is well matched to the GOODS F435W depth, even though the overall number density of sources decreases significantly to shorter wavelengths as already shown visually in Figure \ref{fig:imageExamples}.  }
	\label{fig:diffBandSD}
\end{figure}

\begin{figure}[tbp]
	\centering
	\includegraphics[width=0.96\linewidth]{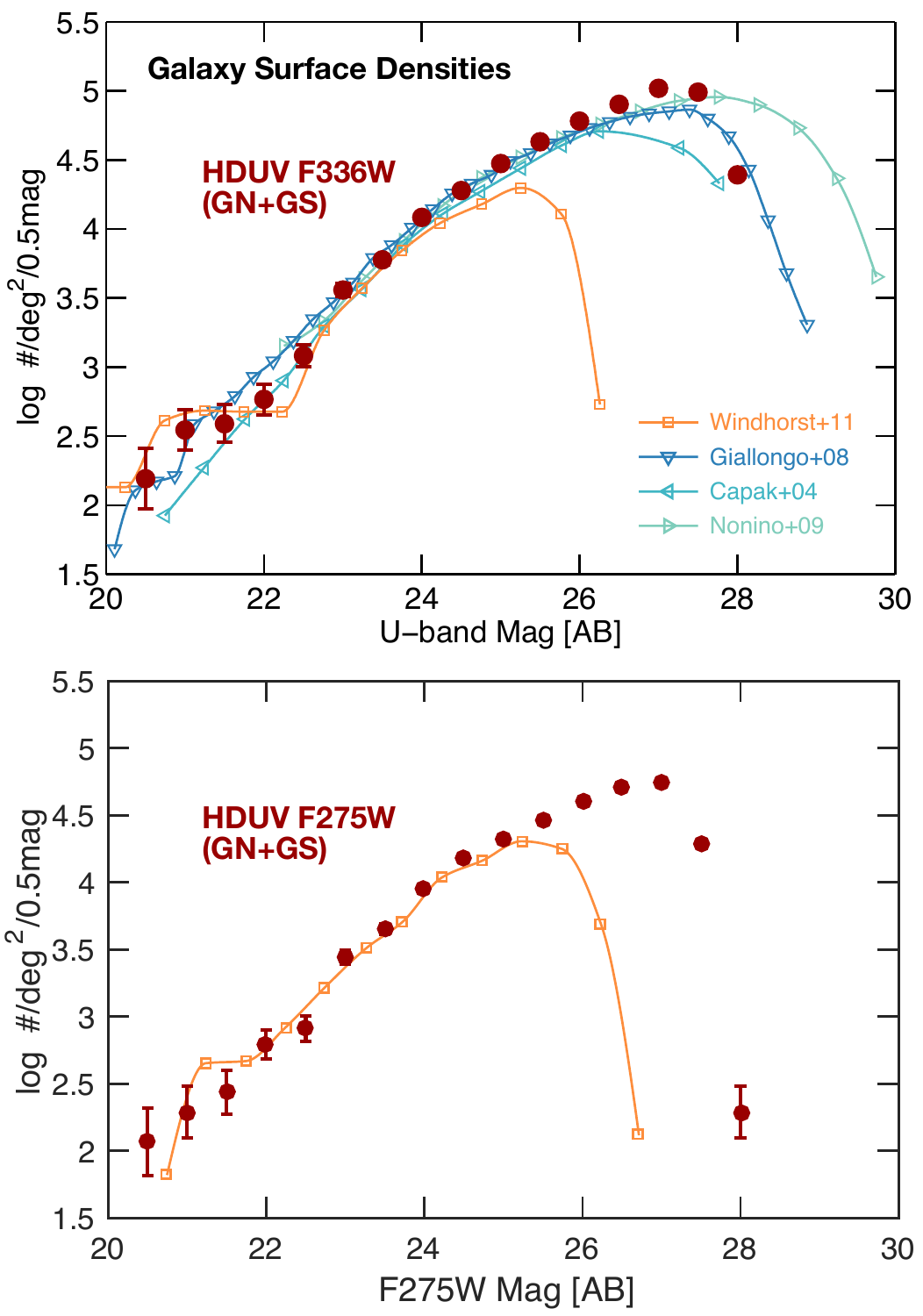}
  \caption{Comparison of the U/F336W-band (top) and F275W (bottom) surface densities of sources measured with different surveys. The surface densities based on the HDUV F336W and F275W filter imaging are shown as dark red dots (with Poisson errorbars). The orange squares and lines represent the F336W/F275W surface density measured in the ERS data \citep{Windhorst11}, while all the other lines correspond to ground-based U-band imaging surveys covering larger areas \citep{Nonino09,Giallongo08,Capak04}. The latter cover slightly longer central wavelengths ($\lambda_c \sim3600-3800$~\AA) compared to $\lambda_c=3350$~\AA\ of F336W, and they all have a much wider PSF ($>8\times$). The curves turn over at faint magnitudes since no completeness corrections have been applied. Magnitudes correspond to total fluxes. }
	\label{fig:ubandSD}
\end{figure}

\subsection{Galaxy Surface Densities and Number Counts}

The $HDUV$ survey image is the first UV image at high spatial resolution to a depth of $>27.5$ mag, which at the same time covers a large area of 100 arcmin$^2$. It is thus interesting to compute the average surface densities of galaxies in the UV compared to other wavelengths. To do this, we detect galaxies in individual bands using SExtractor in single image mode. Figure \ref{fig:diffBandSD} shows the surface densities of sources detected at $>5\sigma$ in three filters, ACS F435W as well as the two HDUV filters. The depth of the HDUV F336W image is well matched to the GOODS F435W data. However, it is clear again that the source density decreases significantly at shorter wavelengths as already highlighted in Figure \ref{fig:imageExamples}. 
Overall, the HDUV survey contains $\sim12,000$ sources that are detected at $>5\sigma$ in F336W and $\sim6,000$ sources in F275W (not including the UVUDF data).

In Figure \ref{fig:ubandSD}, we further compare the F336W/U-band source surface densities seen in different surveys. Previous data include the WFC3/UVIS ERS fields \citep{Windhorst11}, which covered $\sim50$ arcmin$^2$. We also show the surface densities from ground-based U-band surveys such as the 630 arcmin$^2$ very deep VIMOS imaging in the GOODS-South \citep{Nonino09}, a KPNO U-band survey in GOODS-N \citep{Capak04}, and a survey with the Large Binocular Camera \citep{Giallongo08}. Some of these ground-based data reach depths comparable to or even deeper than the HDUV, albeit at $>8\times$ lower resolution. The high spatial resolution is crucial for most of the science questions to be addressed with these UV data, however, including the escape  of ionizing radiation, or the internal stellar population properties of high-redshift galaxies. Additionally, the HDUV extends to shorter wavelengths than can be reached from the ground through the use of the F275W filter. The lower panel of Figure \ref{fig:ubandSD} shows the surface density of this shorter wavelength filter compared to the ERS data.

\begin{figure}[tbp]
	\centering
	\includegraphics[width=0.98\linewidth]{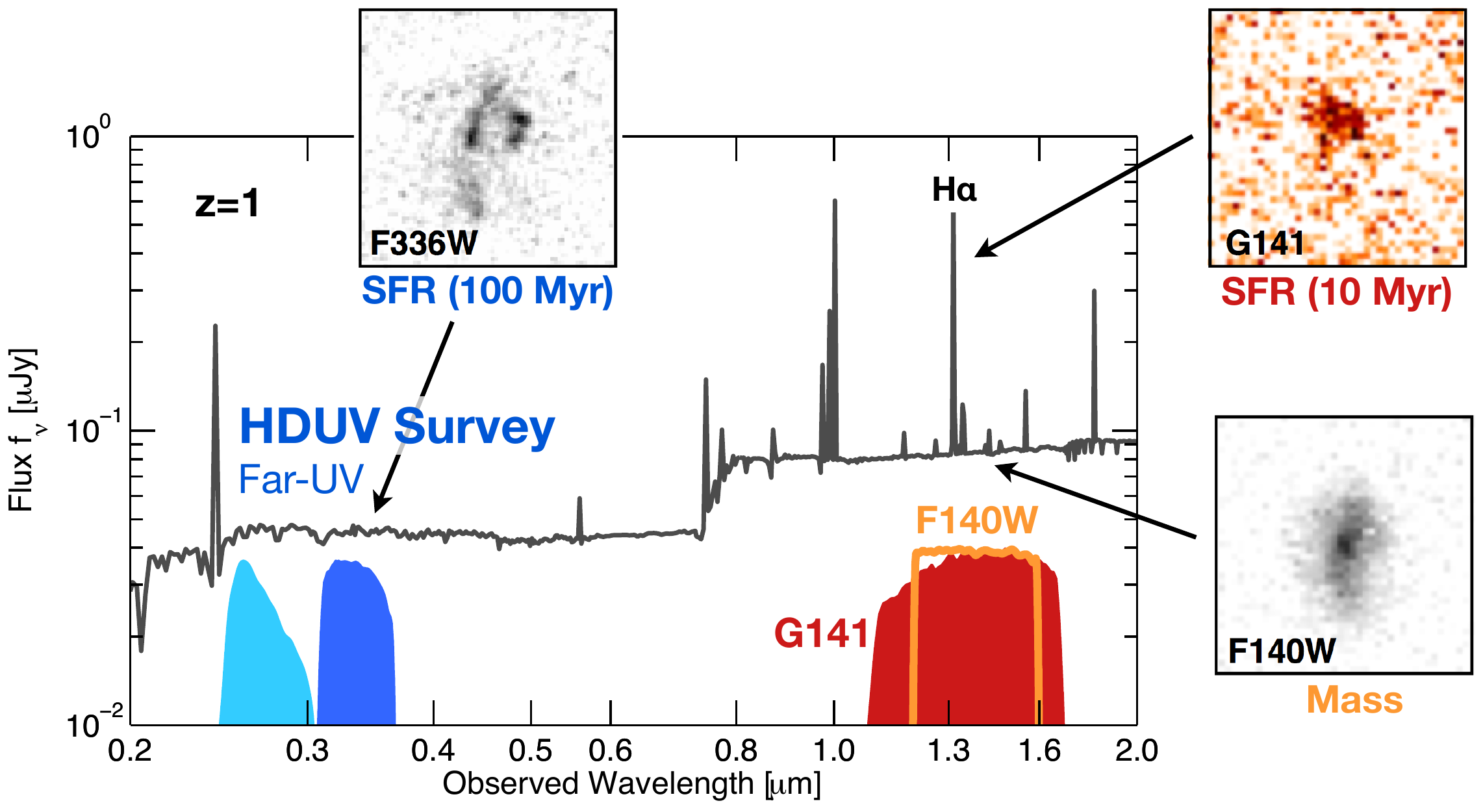}
  \caption{Template spectrum and images of a galaxy at $z\sim1$ highlighting the power of the multi-wavelength data at $HST$ resolution available in the HDUV survey footprint. The HDUV samples the far-UV which is sensitive to star-formation on $\sim100$ Myr timescales, while the WFC3/IR grism provides resolved H$\alpha$ images and the star-formation on $\sim10$ Myr timescales. Finally, the broad-band imaging in the NIR samples the rest-frame optical of galaxies out to $z\sim3$ and is sensitive to the underlying mass distribution. The HDUV thus provides key insights into the internal build-up of galaxies at a spatial resolution of $\sim700-800$ pc out to $z\sim1-3$.}
	\label{fig:viewofz1}
\end{figure}

\section{Summary \& Discussion} 
\label{sec:summary} 

This paper presents the strategy and the data products (v1) of the $HDUV$ survey, a 132 orbit legacy imaging program performed with the WFC3/UVIS camera onboard the $HST$. Two prime extragalactic fields, the GOODS/CANDELS-Deep fields, were imaged in the two filters F275W and F336W, to a depth of 10 and 8 orbits, respectively. When combined with previous UV imaging from the UVUDF survey and from the CANDELS program, this results in  coverage of 100 arcmin$^2$ to a depth of $>27.5$ mag at high spatial resolution ($<0$\farcs1). Overall, the HDUV footprint contains $\sim12'000$ and $\sim6'000$ sources that are detected at $>5\sigma$ in the F336W and F275W filter, respectively.

We show from our data directly that charge transfer inefficiencies of the WFC3/UVIS detector can be adequately corrected for when images are taken with the appropriate postflash levels as recommended by STScI. Even for faint sources, we find only small residual flux biases of typically $<5\%$ after correction (Section \ref{sec:CTEcorr}).

One of the main goals of the HDUV survey was to obtain a complete census of faint, star-forming galaxies at the peak of cosmic star-formation history ($z\sim1-3$). To this end, it was carried out in two extragalactic fields with the most extensive multi-wavelength coverage. When added to the previous ACS and WFC/IR imaging data from the GOODS and CANDELS surveys, the HDUV fields have deep 11-band $HST$ coverage. Additionally, the full HDUV footprint is covered by at least 2-orbit WFC3/IR grism spectroscopy through several surveys, including 3D-HST. These data will thus be of high legacy value for the community. The data release described here includes high-level science products drizzled to the same frame as the 3D-HST image release over the two GOODS fields facilitating their use.

As an example of the power of the different complementary datasets available over these fields, we show the information available for a $z\sim1$ galaxy in Figure \ref{fig:viewofz1}. The HDUV survey provides imaging in the far-UV of such a galaxy, which samples young, massive stars, and hence the star-formation rate on $\sim100$ Myr time-scales. On the other hand, the WFC3/IR grism data provides a view of the star-formation based on the H$\alpha$ emission line, which is sensitive to timescales of $\sim10$ Myr \citep[see e.g.][]{Nelson12,Nelson15}. Finally, the WFC3/IR broad-band imaging samples the rest-frame optical light and is thus sensitive to the galaxies' overall stellar mass distribution.

Quantifying the differences in the SFR distributions on different timescales and the underlying mass distribution (which are clearly substantial for the galaxy shown in Fig \ref{fig:viewofz1}) will be a key ingredient for our understanding of the internal build-up of galaxies. The high spatial resolution of the HDUV data allows for the quantification of galaxies' internal stellar population properties including dust extinction at a resolution of $\sim700-800$ pc out to $z\sim1-3$ (see section \ref{sec:science}).

The calibrated and combined UV data are released to the community as high-level science products through the MAST archive (\dataset[doi:10.17909/T90T2N]{http://dx.doi.org/10.17909/T90T2N}) at \url{https://archive.stsci.edu/prepds/hduv/}. This v1 release includes images at 60 mas pixel scale of both the GOODS-S and GOODS-N field, matched to the same pixel grid as the 3D-HST imaging data release \citep{Skelton14} to facilitate easy multi-wavelength analyses by the community. A separate reduction is drizzled to the pixel grid of the HLF images in GOODS-S, which cover a wider area \citep{Illingworth16}. As no future high-resolution UV imager from space is currently planned, the unique HDUV imaging data will be of crucial legacy value into the $JWST$ era and beyond.

\vspace{0.2cm}
\acknowledgments{
The authors thank the anonymous referee who helped improve this manuscript and encouraged the release of initial catalogs. Further thanks go to Mark Rafelski, Brian Siana, and Harry Teplitz for several very helpful discussions regarding the WFC3/UVIS data acquisition and reduction. The authors also thank Tomer Tal for help during the proposal stage, and Ros Skelton for help with the 3D-HST detection images and SExtractor settings.
   Support for this work was provided by NASA through grant HST-GO-13872 from the Space Telescope Science Institute, which is operated by AURA, Inc., under NASA contract NAS 5-26555. PO further acknowledges support by the Swiss National Science Foundation.
The data presented in this paper were obtained from the Mikulski Archive for Space
Telescopes (MAST). }

Facilities: \facility{HST(ACS/WFC3)}.

\bibliography{MasterBiblio}
\bibliographystyle{apj}

\end{document}